\documentclass[prl,superscriptaddress,preprint,endfloats,nopacs]{revtex4}%preprint type
\usepackage{amsmath, amsthm, amssymb, pifont, wasysym}

\usepackage{graphicx}
\usepackage{bm} % bold math
\usepackage{pifont}
\usepackage{dcolumn} % Align table columns on decimal point
\usepackage{color}
\definecolor{mygray}{gray}{0.5}
\newcommand {\etal}{\textit{et al}.}
\newcommand {\SRO}{Sr$_2$RuO$_4$}
\newcommand {\STO}{SrTiO$_3$}
\newcommand {\RO}{RuO$_2$}
\newcommand {\IO}{IrO$_2$}
\newcommand {\TO}{TiO$_2$}
\newcommand {\MF}{MgF$_2$}
\newcommand {\aone}{$a_{110}$}
\newcommand {\aonebar}{$a_{1\overline{1}0}$}
\newcommand {\aMOone}{$a_{\mathrm{M\mathchar`-O(1)}}$} 
\newcommand {\aMOtwo}{$a_{\mathrm{M\mathchar`-O(2)}}$}
\newcommand {\ttwog}{$t_{2\mathrm{g}}$}
\newcommand {\eg}{$e_{\mathrm{g}}$}

\begin{document}

\title{Superconductivity in Uniquely Strained RuO$_{2}$ Films}

\author{Masaki Uchida}
\email[Author to whom correspondence should be addressed: ]{uchida@ap.t.u-tokyo.ac.jp}
\affiliation{Department of Applied Physics, University of Tokyo, Tokyo 113-8656, Japan}
\affiliation{Quantum-Phase Electronics Center (QPEC), University of Tokyo, Tokyo 113-8656, Japan}
\affiliation{PRESTO, Japan Science and Technology Agency (JST), Tokyo 102-0076, Japan}
\author{Takuya Nomoto}
\affiliation{Department of Applied Physics, University of Tokyo, Tokyo 113-8656, Japan}
\author{Maki Musashi}
\affiliation{Department of Applied Physics, University of Tokyo, Tokyo 113-8656, Japan}
\affiliation{Quantum-Phase Electronics Center (QPEC), University of Tokyo, Tokyo 113-8656, Japan}
\author{Ryotaro Arita}
\affiliation{Department of Applied Physics, University of Tokyo, Tokyo 113-8656, Japan}
\affiliation{RIKEN Center for Emergent Matter Science (CEMS), Wako 351-0198, Japan}
\author{Masashi Kawasaki}
\affiliation{Department of Applied Physics, University of Tokyo, Tokyo 113-8656, Japan}
\affiliation{Quantum-Phase Electronics Center (QPEC), University of Tokyo, Tokyo 113-8656, Japan}
\affiliation{RIKEN Center for Emergent Matter Science (CEMS), Wako 351-0198, Japan}

%%%%%%%%%%%%%%%%%%%%%%
\begin{abstract}
We report strain engineering of superconductivity in {\RO} singlecrystalline films, which are epitaxially grown on rutile {\TO} and {\MF} substrates with various crystal orientations. Systematic mappings between the superconducting transition temperature and the lattice parameters reveal that shortening of specific ruthenium-oxygen bonds is a common feature among the superconducting {\RO} films. Ab initio calculations of electronic and phononic structures for the strained {\RO} films suggest the importance of soft phonon modes for emergence of the superconductivity. The findings indicate that simple transition metal oxides such as with the rutile structure may be suitable for further exploring superconductivity by controlling phonon modes through the epitaxial strain.
\end{abstract}
\maketitle
%%%%%%%%%%%%%%%%%%%%%%

Transition metal oxides represented by cuprates and ruthenates have guided us to better understanding of unconventional superconductivity originating in the strong electron correlation \cite{oxideSC}. In some other oxides such as titanates, on the other hand, it has been thought that lattice vibrations or phonons play a more dominant role for pairing electrons in the superconducting state. In this context, binary oxide superconductors including Ti$_{n}$O$_{2n-1}$ ($n=1$--$4$) \cite{Mono_TiO, Mono_TiOandNbO, Mono_TiO_film, Mono_Ti4O7andTi3O5_film, Mono_Ti3O5_film2, Mono_Ti2O3_film}, NbO \cite{Mono_TiOandNbO}, SnO \cite{Mono_SnO}, and LaO \cite{Mono_LaO_film} are intriguing systems. Many of them have been realized only in epitaxial films and their superconducting mechanisms are still elusive. In these films, epitaxial strain which directly tunes lattice parameters is expected to be useful for designing superconductivity by controlling phonon modes, electron correlation, and so on, as recently demonstrated in {\STO} thin films \cite{STO1, STO2}.

{\RO} with the rutile structure is well known to be a highly conducting binary oxide \cite{RuO2_transport1, RuO2_transport2}. {\RO} finds many engineering applications in electrodes, thermometers, and also catalysts, and thin films, mostly polycrystalline films, have been prepared for such purposes by various growth methods \cite{film_sputtering1_poly, film_sputtering_other, film_sputtering2_poly, film_CVD1, film_CVD_other1, film_CVD_other2, film_PLD_other3, film_PLD1, film_PLD2, film_PLD_other1, film_MBE}. In recent years, on the other hand, {\RO} has attracted renewed attention as a high-temperature antiferromagnet \cite{RuO2_afm1, RuO2_afm2} and a possible topological nodal line semimetal \cite{RuO2_TNLS, TopoOxidereview}, demanding reexamination of its electronic transport in the ground state. It has been also reported that superconductivity appears in {\RO} thin films grown on a rutile {\TO} substrate \cite{APS}. Motivated by this, here we systematically investigate epitaxial strain effect on superconductivity in {\RO} films.

{\RO} thin films were grown on single-crystalline rutile {\TO} and {\MF} substrates with various crystal orientations in an oxide molecular beam epitaxy system \cite{UchidaMBE, Hc2_thinfilm, JJ_thinfilm}, referring to molecular beam epitaxy of {\IO} with the same rutile structure and similar volatile binary phases \cite{IrO2_MBE1, IrO2_MBE2}. 3N5 Ru elemental flux was supplied from an electron beam evaporator. Optimized growth was performed at a substrate temperature of 300 $^{\circ}$C, regulated with a semiconductor-laser heating system, and with flowing pure $\mathrm{O}_{3}$ with a pressure of $6\times 10^{-7}$ Torr, supplied from a Meidensha Co. ozone generator. The film thickness was adjusted in the range of 26 to 32 nm to effectively apply large epitaxial strain. Longitudinal resistivity was measured with a standard four-probe method in a Quantum Design PPMS cryostat equipped with a 9 T superconducting magnet and a 3He refrigerator. Density functional theory calculations were performed by using Quantum Espresso package \cite{calc1_QE}. The exchange correlation functional proposed by Perdew {\etal} \cite{calc2_PBE} and pseudopotentials by Garrity {\etal} \cite{calc3_GBRD} were used in the calculations. Phonon band structures were obtained by using density functional perturbation theory \cite{calc4_DEPT}. The $12 \times 12 \times 12$ $k$-points and $3 \times 3 \times 3$ $q$-points were used for the electronic structure calculations and the dynamical matrix calculations, respectively.

\begin{figure}
\begin{center}
\includegraphics*[width=13.5cm]{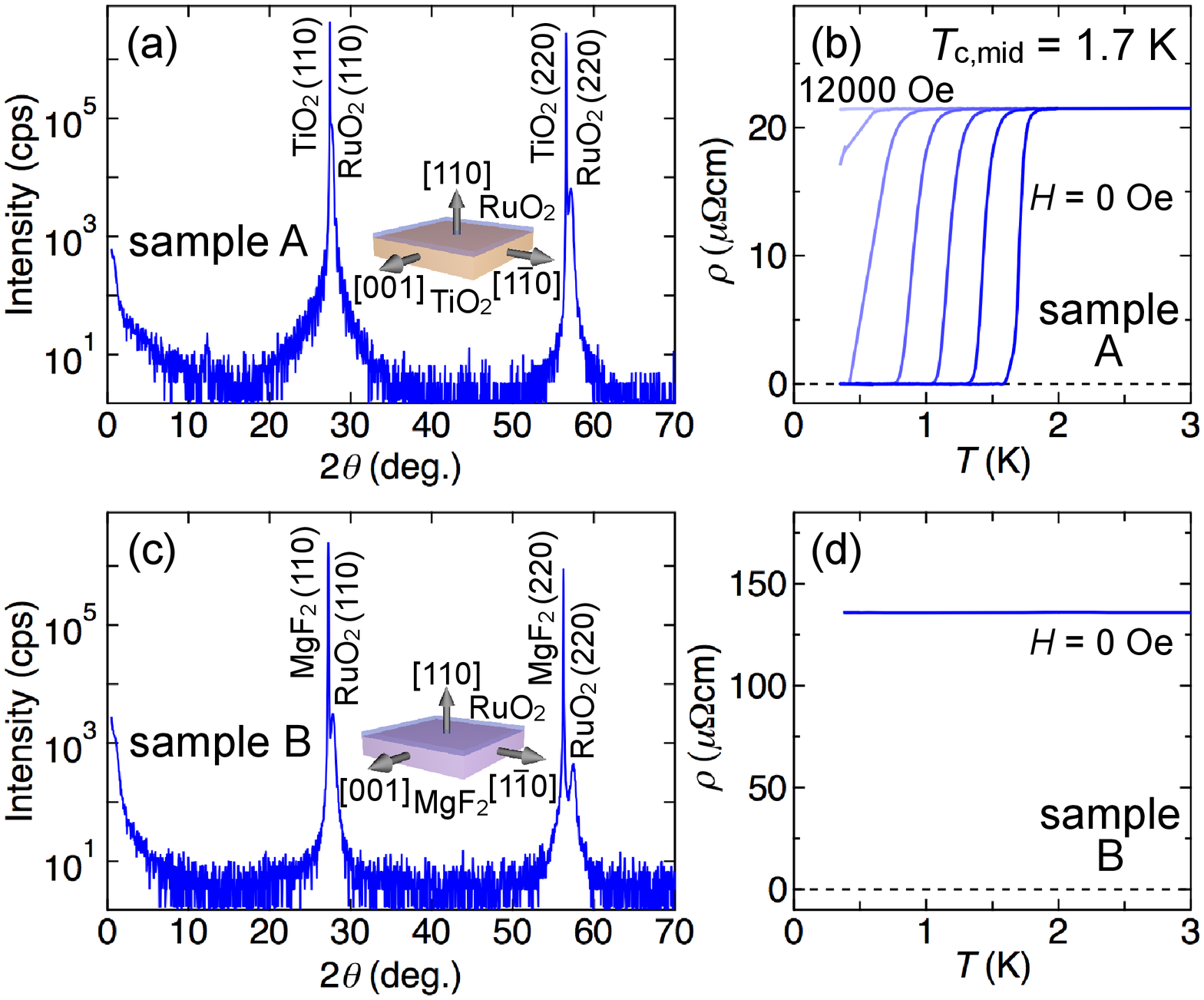}
\caption{
Superconducting and non-superconducting {\RO} thin films. (a) XRD $\theta$--2$\theta$ scan of sample A, which is a {\RO} film grown on the (110)-oriented {\TO} substrate. (b) Temperature dependence of the in-plane resistivity for sample A, measured with applying a magnetic field parallel to the out-of-plane direction at intervals of 2000 Oe. It shows a clear superconducting transition with a midpoint temperature of $T_{\mathrm{c, mid}} =1.7$ K. (c) XRD scan of sample B, a {\RO} film grown on the (110)-oriented {\MF} substrate. (d) Low-temperature resistivity of sample B down to 0.4 K.
}
\label{fig1}
\end{center}
\end{figure}

Figure 1(a) shows x-ray diffraction (XRD) $\theta$-2$\theta$ scan of a {\RO} thin film grown on the (110)-oriented {\TO} substrate (sample A). It shows only sharp ($ll$0) {\RO} peaks ($l$:  integer) nearby the substrate ones, indicating that single-crystalline {\RO} is epitaxially grown on the substrate with the same rutile structure. As confirmed in Fig. 1(b), longitudinal resistivity of sample A begins to drop at $T_{\mathrm{c,onset}}=1.8$ K, decreases by half at $T_{\mathrm{c, mid}} =1.7$ K, and then becomes zero at $T_{\mathrm{c,zero}}=1.6$ K. With increasing the out-of-plane magnetic field up to 12000 Oe, the superconducting transition gradually shifts to lower temperatures and eventually disappears. This is in contrast to the other ruthenate superconductor {\SRO} \cite{SRO}, in which the out-of-plane upper critical field is much lower (750 Oe in bulks and 2200 Oe in thin films) while the transition temperature is comparable \cite{Hc2_yonezawafirst1, Hc2_thinfilm}. 

Another {\RO} film on the (110)-oriented {\MF} substrate (sample B) is also epitaxially grown in the single-crystalline form, as shown in Fig. 1(c). In Fig. 1(d), on the other hand, sample B does not exhibit the superconducting transition down to 0.4 K. This suggests the importance of elaborate strain engineering for the emergence of superconductivity in {\RO}. Possibly due to misfit dislocations stemming from the lattice mismatched heterointerfaces, residual resistivity of these samples is much higher than values of about 0.05 to 2 $\mu\Omega$cm reported in {\RO} bulks \cite{RuO2_transport1, RuO2_transport2}. On the other hand, there is no definite correlation between the residual resistivity and the emergence of superconductivity among all the samples discussed below \cite{supplement}.

\begin{figure}
\begin{center}
\includegraphics*[width=12.5cm]{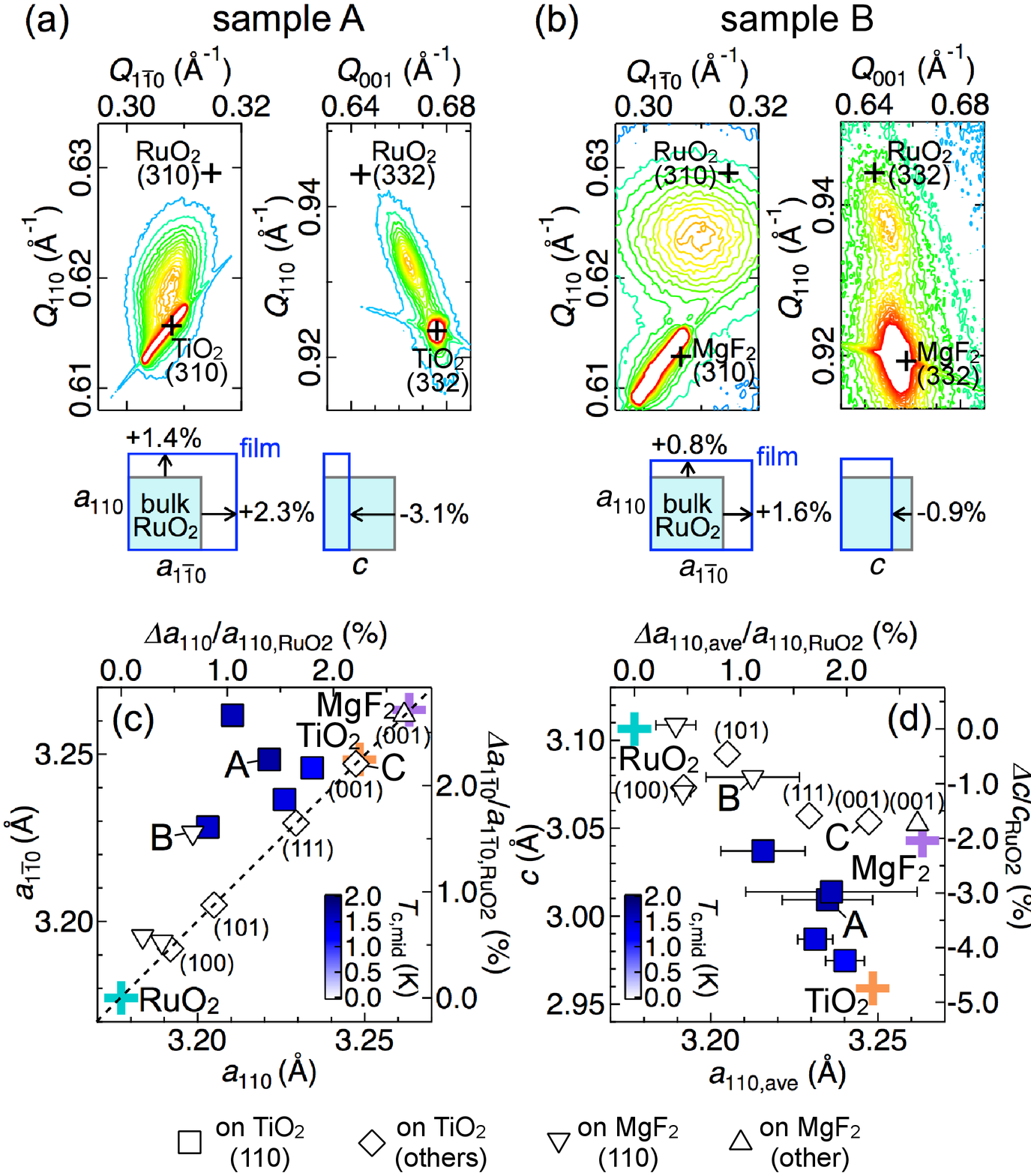}
\caption{
Superconductivity in uniquely strained {\RO} films. (a) and (b) XRD reciprocal space mappings in samples A and B, taken for asymmetric reflections both in the (1$\overline{1}$0) and (001) in-plane directions. A cross denotes peak positions calculated from bulk lattice parameters of {\RO}, {\TO}, and {\MF}. Changes in the {\RO} lattice parameters from the bulk values are also schematically illustrated. (c) Mapping of the lattice parameters {\aone} vs {\aonebar} measured for all the samples grown on {\TO} or {\MF} substrates with different crystal orientations as shown near the symbols. This includes sample C, grown on the (001)-oriented {\TO} substrate. The symbols are colored by the midpoint superconducting transition temperature. (d) Mapping of $a_{110, {\mathrm{ave}}}$ vs $c$ also colored by the superconducting transition temperature. $a_{110, {\mathrm{ave}}}$ means the average of $a_{110}$ and $a_{1\overline{1}0}$, which are shown as both ends of the horizontal bar appended for the case of the anisotropic strain. A cross indicates bulk lattice parameters of {\RO}, {\TO}, and {\MF}.
}
\label{fig2}
\end{center}
\end{figure}

Reciprocal space mappings in Figs. 2(a) and 2(b) demonstrate how the epitaxial strain affects film lattice parameters {\aone}, {\aonebar}, and $c$, which are along three directions (($110$), ($1\overline{1}0$), and ($001$)) orthogonal to each other. In sample A, the film lattice is coherently grown and fully strained by $+2.3$\% in complete matching with the substrate lattice along the in-plane ($1\overline{1}0$) direction, while it is partially strained not in matching with the substrate one along the other in-plane (001) direction. In sample B, on the other hand, the lattice is partially strained both for the ($1\overline{1}0$) and (001) directions by $+1.6$\% and $-0.9$\%. Namely, while {\aone} and {\aonebar} are anisotropically extended in both the (110)-oriented films, $c$ is greatly shortened in sample A compared to sample B, reflecting the large $c$-axis mismatch of $-4.7$\% between {\TO} and {\RO}. 

Figures 2(c) and 2(d) summarize the three lattice parameters {\aone}, {\aonebar}, and $c$ measured for all the samples including other orientation films, with comparing their changes to the {\RO}, {\TO}, and {\MF} bulks. For example, a film grown on the (001)-oriented {\TO} substrate (sample C) is fully strained along the in-plane (110) and ($1\overline{1}0$) directions, as confirmed in Fig. 2(c). This is also the case on the (001)-oriented {\MF} substrate. Actually, except the (110)-oriented films such as samples A and B, {\aone} and {\aonebar} are isotropically extended in all the other films. As well as such isotropically strained films, no superconductivity appears even in the (110)-oriented films on {\MF}, although the {\aone} and {\aonebar} values especially of sample B are nearly equal to the ones of the (110)-oriented superconducting films on {\TO}. Rather, correlation between the superconductivity and the epitaxial strain is clearly visualized in the mapping for the $c$-axis change in Fig. 2(d). The superconductivity emerges only in the (110)-oriented {\RO} films grown on {\TO}, where $c$ is shortened by $2$\% or much more, unlike the other films.

\begin{figure}
\begin{center}
\includegraphics*[width=13.5cm]{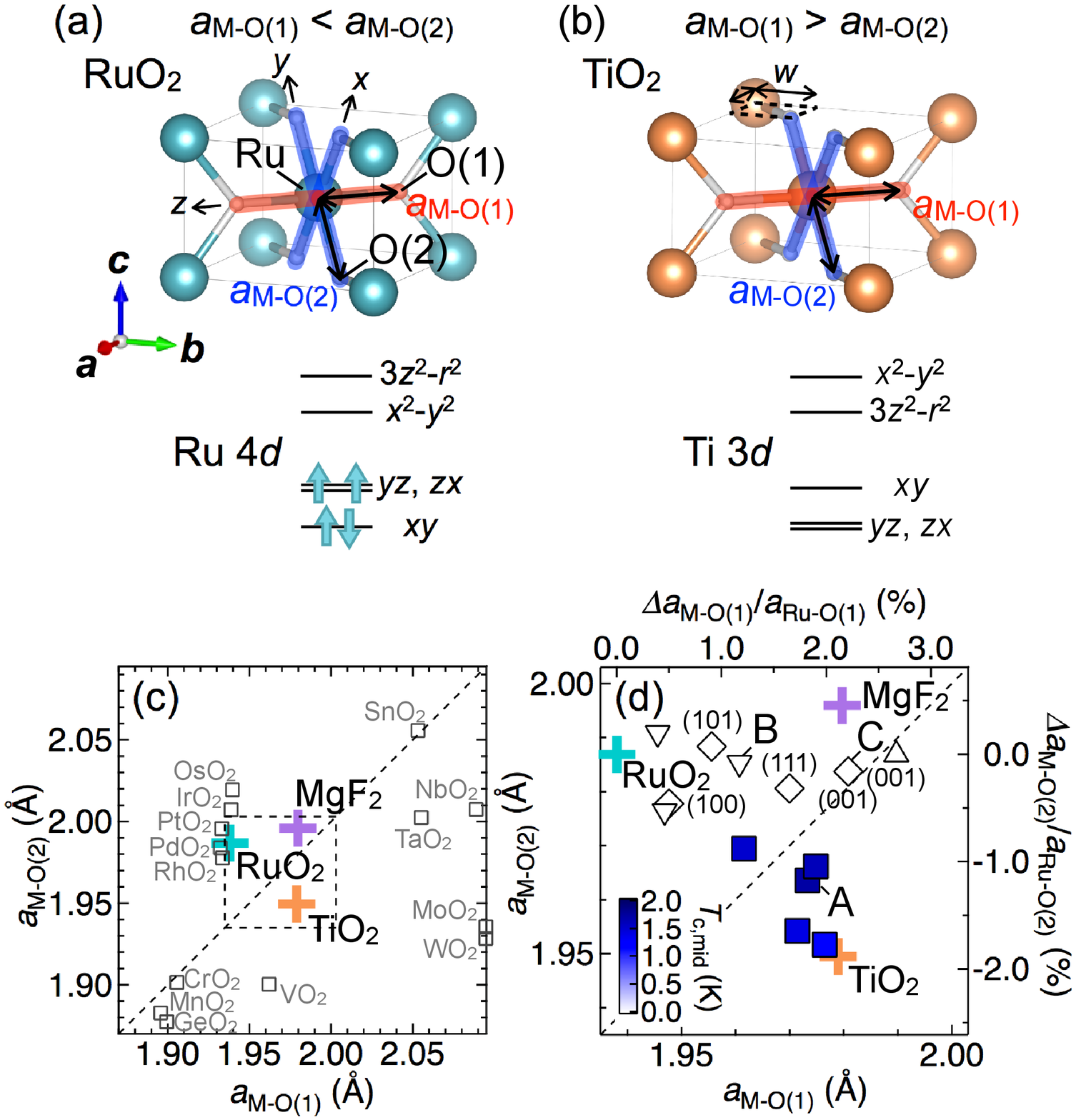}
\caption{
Two types of the rutile structure. (a) Rutile crystal structure of bulk {\RO}, composed of short M-O(1) and long M-O(2) bonds. Simple configuration of the $4d^4$ electrons determined by the resultant crystal field splitting is also shown. (b) Case of {\TO}, where the M-O(1) (M-O(2)) bonds are inversely extended (shortened) and the energy splittings within the {\ttwog} and {\eg} levels are also reversed. (c) Mapping of the two bond lengths {\aMOone} vs {\aMOtwo}, calculated from the lattice parameters $a$ and $c$ and the Wyckoff position coordinate $w$ for many rutile compounds. (d) {\aMOone} vs {\aMOtwo} similarly estimated for all the {\RO} film samples, plotted on magnified area corresponding to the dashed box in (c). Here the average is plotted for anisotropic cases where $a_{110}$ and $a_{1\overline{1}0}$ have different values. The symbols are colored by the superconducting transition temperature as in Figs. 2(c) and 2(d).
}
\label{fig3}
\end{center}
\end{figure}

Rutile oxides can be roughly categorized into two types, by the magnitude relation between two metal(M)-oxygen(O) bond lengths {\aMOone} and {\aMOtwo} as illustrated in Figs. 3(a) and 3(b). In {\RO} bulk with M-O(2) longer than M-O(1), the $d_{xy}$ state has slightly lower energy and is fully occupied by the 4$d$ electrons, while the $d_{yz}$/$d_{zx}$ states are half-filled. In {\TO} bulk, on the other hand, M-O(1) is longer than M-O(2), and thus the energy splitting between the $d_{xy}$ and $d_{yz}$/$d_{zx}$ states is expected to be reversed. {\aMOone} and {\aMOtwo} can be calculated from the lattice parameters $a$ and $c$ and the Wyckoff position coordinate $w$ using the following relations.
\begin{center}
\begin{eqnarray}
a_{\mathrm{M\mathchar`-O(1)}}&=& \sqrt{2}wa\\
a_{\mathrm{M\mathchar`-O(2)}} &=& \sqrt{(\sqrt{2}(0.5-w)a)^2+(c/2)^2}
\end{eqnarray}
\end{center}
Here $w=0.305$ is used for film samples because $w$ is almost independent of the compounds and within the range between 0.3045 and 0.3065. As confirmed in Fig. 3(c), {\RO} and {\TO} are respectively located at the two regions of $a_{\mathrm{M\mathchar`-O(1)}} < a_{\mathrm{M\mathchar`-O(2)}}$ and $a_{\mathrm{M\mathchar`-O(1)}} > a_{\mathrm{M\mathchar`-O(2)}}$, while averages of {\aMOone} and {\aMOtwo} are almost the same. {\MF}, which has near-{\TO} {\aMOone} and near-{\RO} {\aMOtwo}, is rather close to the region boundary. Figure 3(d) reveals an important trend of {\aMOone} vs {\aMOtwo}, which distinguishes the superconducting and non-superconducting {\RO} films. Namely, in the superconducting films (e.g. sample A), {\aMOtwo} substantially decreases with approaching the {\TO} bulk value, in addition to the increase of {\aMOone}. On the other hand, only the increase of {\aMOone} is confirmed in the non-superconducting films (e.g. samples B and C), whose parameters are distributed between the {\RO} and {\MF} bulks. This map clearly indicates that the shortening of the M-O(2) bonds is essential for emergence of the superconductivity.

\begin{figure}
\begin{center}
\includegraphics*[width=13.5cm]{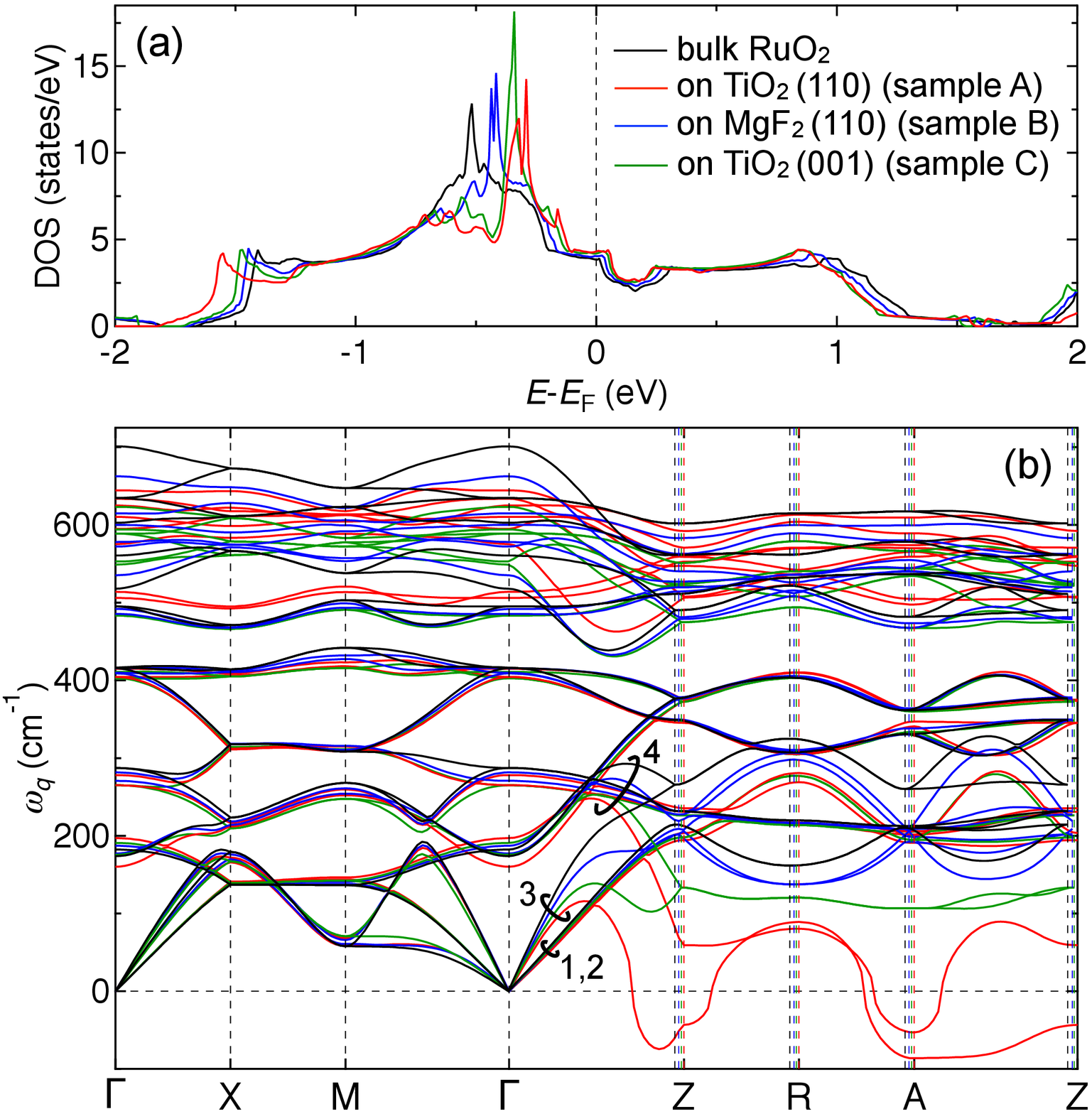}
\caption{
Electronic and phononic changes in the strained {\RO} films. (a) Electron density of states calculated for {\RO} bulk and thin films (samples A, B, and C). (b) Phonon band structures calculated for the same set of samples. In superconducting sample A, acoustic and optical phonon modes oscillating along the $c$ direction (modes 3 and 4) show clear softening especially around the Z and A points.
}
\label{fig4}
\end{center}
\end{figure}

Next we examine strain effects on fundamental electronic and phononic properties in {\RO}. Figure 4(a) compares electron density of states between the {\RO} bulk and strained thin films (samples A, B, and C). The density of states ranging between $E-E_{\mathrm{F}} \sim -1.5$ and $1.2$ eV is mainly from the Ru {\ttwog} bands and the bulk band structure is consistent with previous calculations \cite{RuO2_calc1}. In the film samples with the extended M-O(1) and shortened M-O(2) bonds or even only with the extended M-O(1) bonds, it is expected that the $d_{xy}$ band is relatively shifted to higher energies. However, a peak at $E-E_{\mathrm{F}} \sim -0.5$ eV ascribed to the $d_{xy}$ band, for example, is shifted only about $0.2$ eV even in superconducting sample A, which is small compared to the band width. In particular, the calculated density of states at the Fermi level is not substantially different among the superconducting and non-superconducting samples.

In contrast, phonon dispersion relations are largely modulated by the lattice parameter change. As shown in Fig. 4(b), two specific phonon modes in superconducting sample A exhibit negative frequencies indicating dynamical instability. Nearly degenerate modes originally with the lowest energies along the $\Gamma$-Z line, which are the acoustic modes oscillating on the $a$-$b$ plane (modes 1 and 2), remain almost unchanged. On the other hand, phonon softening occurs in one with the second lowest energies, which is the acoustic mode oscillating along the $c$ direction (mode 3). The similar softening also occurs in the optical mode oscillating along the $c$ direction (mode 4). With approaching the Z point, the $c$-direction modes 3 and 4 are more complicatedly hybridized with optical modes of O atoms oscillating on the $a$-$b$ plane. These calculations suggest the possibility that pairing of the Ru 4$d$ electrons is mainly mediated by the soft phonon modes induced by the shortening of the $c$-axis or the M-O(2) bonds. Actually, Eliashberg spectral function $\alpha^2F$ calculated for the same set of samples shows that the soft phonon modes give a large spectral weight in superconducting sample A, suggesting that the low-frequency part of the soft phonon modes mainly contributes to the superconductivity (for details see the Supplemental Material \cite{supplement}).

In summary, we have studied epitaxial strain effect on superconductivity in {\RO} thin films, by combining x-ray diffraction characterization, low-temperature transport measurement, and first-principles calculations for various types of strained films. In particular, comparison between the (110)-oriented films grown on the {\TO} and {\MF} substrates has clarified that shortening of the $c$-axis or the M-O(2) bonds is essential for emergence of the superconductivity. The theoretical calculations have demonstrated that softening of the two phonon modes oscillating along the $c$ direction is dramatically induced by the lattice parameter change. This study suggests that the epitaxial strain will become a powerful tool for directly tuning phonon modes and their mediated superconductivity especially in simple transition metal oxides such as with the rutile structure, in addition to mechanical pressure \cite{softeningSC0_pressure} and chemical substitution \cite{softeningSC1_chemical, softeningSC2_chemical}. On the other hand, the superconductivity in this system may be affected also by electron correlation or antiferromagnetic ordering, which are not included in the present calculations. We hope that our study will trigger further exploration of superconductivity in the simple transition metal oxides by tuning phonon dispersion relations and/or electron phonon interaction through the epitaxial strain.

We would like to thank Y. Tokura and Y. Motome for fruitful discussions and K. S. Takahashi for experimental advice. This work was supported by Grant-in-Aids for Scientific Research on Innovative Areas No. 19H05825, Scientific Research (S) No. 16H06345, Scientific Research (B) No. JP18H01866, and Early-Career Scientists JP19K14654 from MEXT, Japan and by JST PRESTO No. JPMJPR18L2 and CREST Grant No. JPMJCR16F1, Japan.

\end{document}

\title{Supplemental Material for\\
Superconductivity in Uniquely Strained RuO$_{2}$ Films}

\author{Masaki Uchida}
\email[Author to whom correspondence should be addressed: ]{uchida@ap.t.u-tokyo.ac.jp}
\affiliation{Department of Applied Physics, University of Tokyo, Tokyo 113-8656, Japan}
\affiliation{Quantum-Phase Electronics Center (QPEC), University of Tokyo, Tokyo 113-8656, Japan}
\affiliation{PRESTO, Japan Science and Technology Agency (JST), Tokyo 102-0076, Japan}
\author{Takuya Nomoto}
\affiliation{Department of Applied Physics, University of Tokyo, Tokyo 113-8656, Japan}
\author{Maki Musashi}
\affiliation{Department of Applied Physics, University of Tokyo, Tokyo 113-8656, Japan}
\affiliation{Quantum-Phase Electronics Center (QPEC), University of Tokyo, Tokyo 113-8656, Japan}
\author{Ryotaro Arita}
\affiliation{Department of Applied Physics, University of Tokyo, Tokyo 113-8656, Japan}
\affiliation{RIKEN Center for Emergent Matter Science (CEMS), Wako 351-0198, Japan}
\author{Masashi Kawasaki}
\affiliation{Department of Applied Physics, University of Tokyo, Tokyo 113-8656, Japan}
\affiliation{Quantum-Phase Electronics Center (QPEC), University of Tokyo, Tokyo 113-8656, Japan}
\affiliation{RIKEN Center for Emergent Matter Science (CEMS), Wako 351-0198, Japan}

%%%%%%%%%%%%%%%%%%%%%%%%%%%%%%%
%
\maketitle
\section{Residual resistivity}

Table S1 shows superconducting transition temperature and residual resistivity measured for typical {\RO} films in each combination of substrates and orientations. Some {\RO} films with residual resistivity comparable to or even lower than the superconducting one grown on the (110)-oriented {\TO} substrate do not show superconductivity, indicating that there is no definite correlation between the residual resistivity and the emergence of superconductivity.

\begin{table}[h]
\caption{Midpoint superconducting transition temperature $T_{\mathrm{c, mid}}$ and residual resistivity $\rho_{2\mathrm{K}}$ at 2 K measured for typical {\RO} films, grown on {\TO} and {\MF} substrates with various crystal orientations. \vspace{3.0mm}}
  \begin{tabular}{ccccc} \hline
    \multicolumn{1}{c}{\;substrate\;} & \multicolumn{1}{c}{\;orientation\;} & \multicolumn{1}{c}{\;$T_{\mathrm{c, mid}}$ (K)\;} & \multicolumn{1}{c}{\;$\rho_{2\mathrm{K}}$ ($\mu \Omega$cm)\;} & \multicolumn{1}{c}{\;name\;}\\ \hline\hline
    \multicolumn{1}{c}{} & \multicolumn{1}{c}{\;(110)\;} & \multicolumn{1}{c}{\;1.7\;} & \multicolumn{1}{c}{\;21\;} & \multicolumn{1}{c}{\;sample A\;} \\ 
    \multicolumn{1}{c}{} & \multicolumn{1}{c}{\;(001)\;} & \multicolumn{1}{c}{\;0\;} & \multicolumn{1}{c}{\;72\;} & \multicolumn{1}{c}{\;sample C\;}\\ 
     {\TO} & \;(100)\; & \;0\; & \;35\; & \;--\;\\ 
      & \;(101)\; & \;0\; & \;18\; & \;--\;\\ 
     & \;(111)\; & \;0\; & \;38\; & \;--\;\\ \hline
     \multirow{2}{*}{{\MF}} & \;(110)\; & \;0\; & \;136\; & \;sample B\;\\ 
      & \;(001)\; & \;0\; & \;178\; & \;--\;\\ \hline
  \end{tabular}
\end{table}

\section{Film thickness dependence}

Figure S1 shows thickness dependence of the midpoint superconducting transition temperature taken for {\RO} films grown on the (110)-oriented {\TO} substrate. In thicker films than 40 nm, the large epitaxial strain of about 3\% or more cannot be sustained. As shown in the reciprocal space mappings in Fig. S2(a), the $c$-axis is compressed only by 1.9\% in a 48 nm-thick {\RO} film. This $c$-axis compression is understood not enough to induce the superconductivity in {\RO}, as indicated by the red open square in Fig. S2(b). With reducing the film thickness below 20 nm, on the other hand, the superconducting transition is also rapidly suppressed. 20 nm is a bit thick to ascribe the suppression of superconductivity to simple interface effects, and this suppression might be related to other exotic effects such as low-dimensional effect on the phonon softening. Therefore, this thickness dependence is interesting in itself and deserves a future study.

\begin{figure}
\begin{center}
\includegraphics*[width=8.0cm]{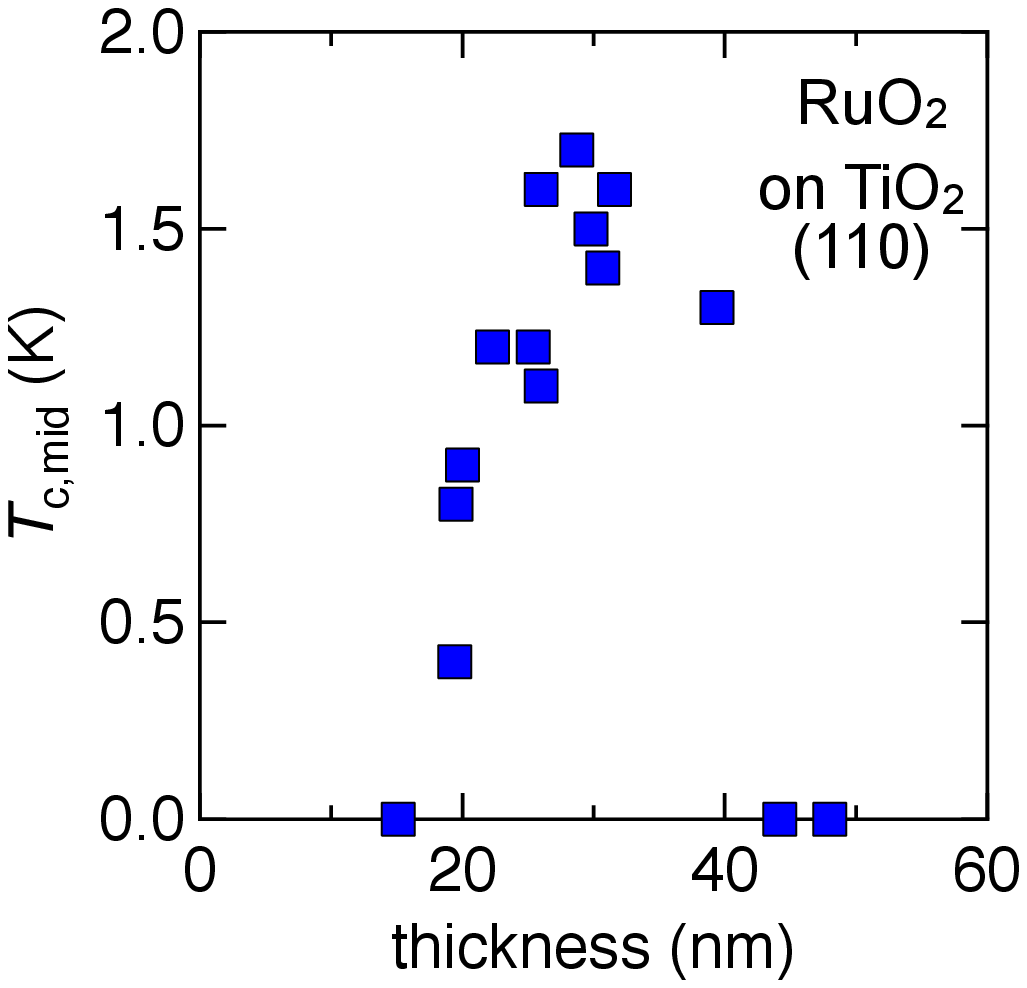}
\caption{
Thickness dependence of the midpoint superconducting transition temperature taken for {\RO} films grown on the (110)-oriented {\TO} substrate.
}
\label{figS2}
\end{center}
\end{figure}

\begin{figure}
\begin{center}
\includegraphics*[width=12.5cm]{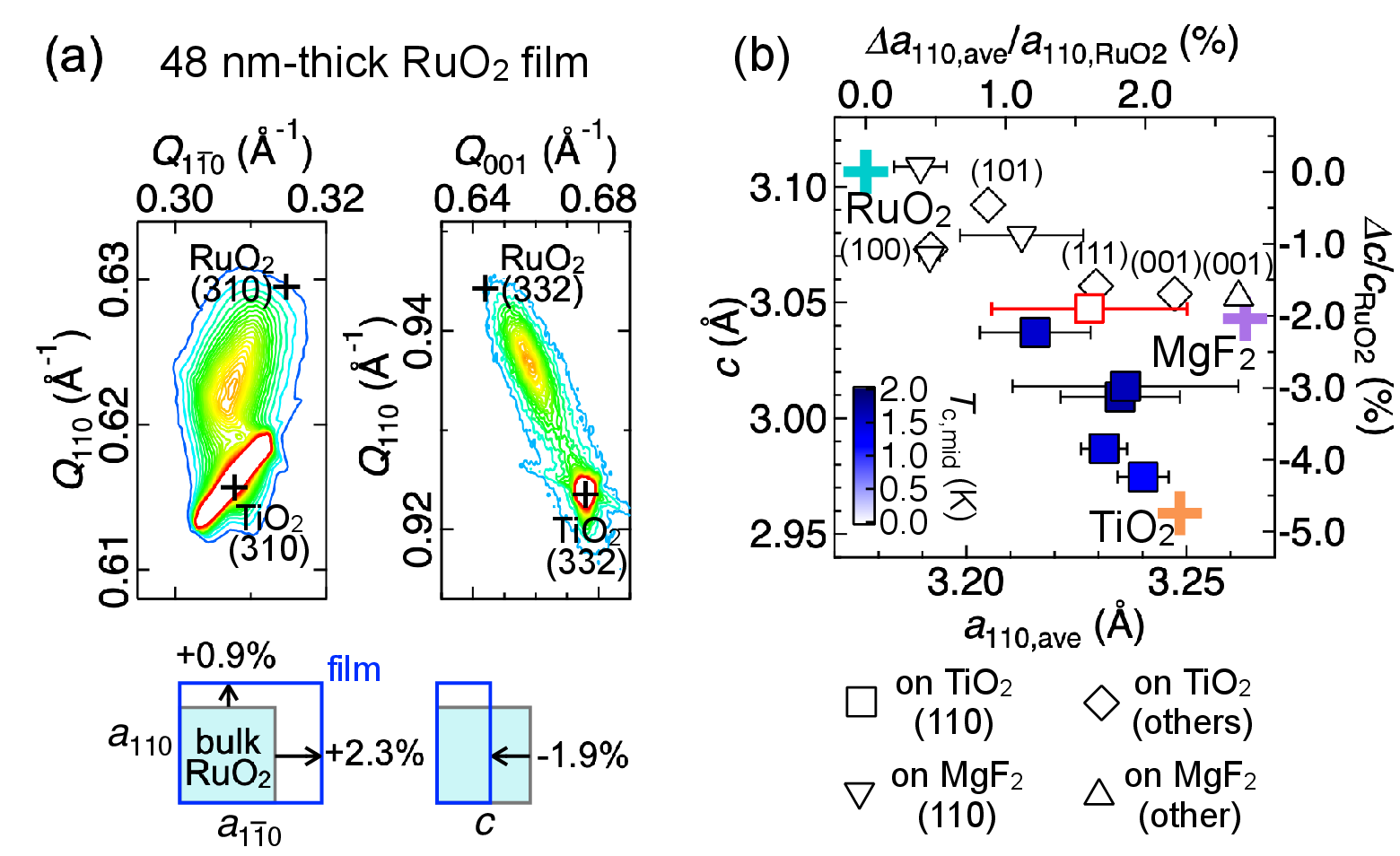}
\caption{
(a) XRD reciprocal space mappings taken for a 48 nm-thick {\RO} film grown on the (110)-oriented {\TO} substrate. A cross denotes peak positions calculated from bulk lattice parameters of {\RO} and {\TO}. Changes in the {\RO} lattice parameters from the bulk values are also schematically illustrated. (b) Plot of the 48-nm thick film data (red square) on the mapping of the lattice parameter changes colored by the superconducting transition temperature in Fig. 2(d).
}
\label{figS2}
\end{center}
\end{figure}

\clearpage

\section{First-principles calculations of the Eliashberg spectral function}
As confirmed in Fig. 4(b), phonon softening occurs around the Z and A points with shortening of the $c$-axis or the M-O(2) bonds in {\RO}. In addition to the change in phonon band structures, we also calculate the Eliashberg spectral function $\alpha^2F(\omega)$ for the same set of samples, as shown in Fig. S3. Along with the phonon softening, a peak seen at 160 cm$^{-1}$ in the bulk grows with shifting to low frequencies, and then shows a large weight in superconducting sample A. This suggests that the low-frequency part of the soft phonon modes, not the original acoustic phonon modes, contributes to the superconductivity.

\begin{figure}
\begin{center}
\includegraphics*[width=12.5cm]{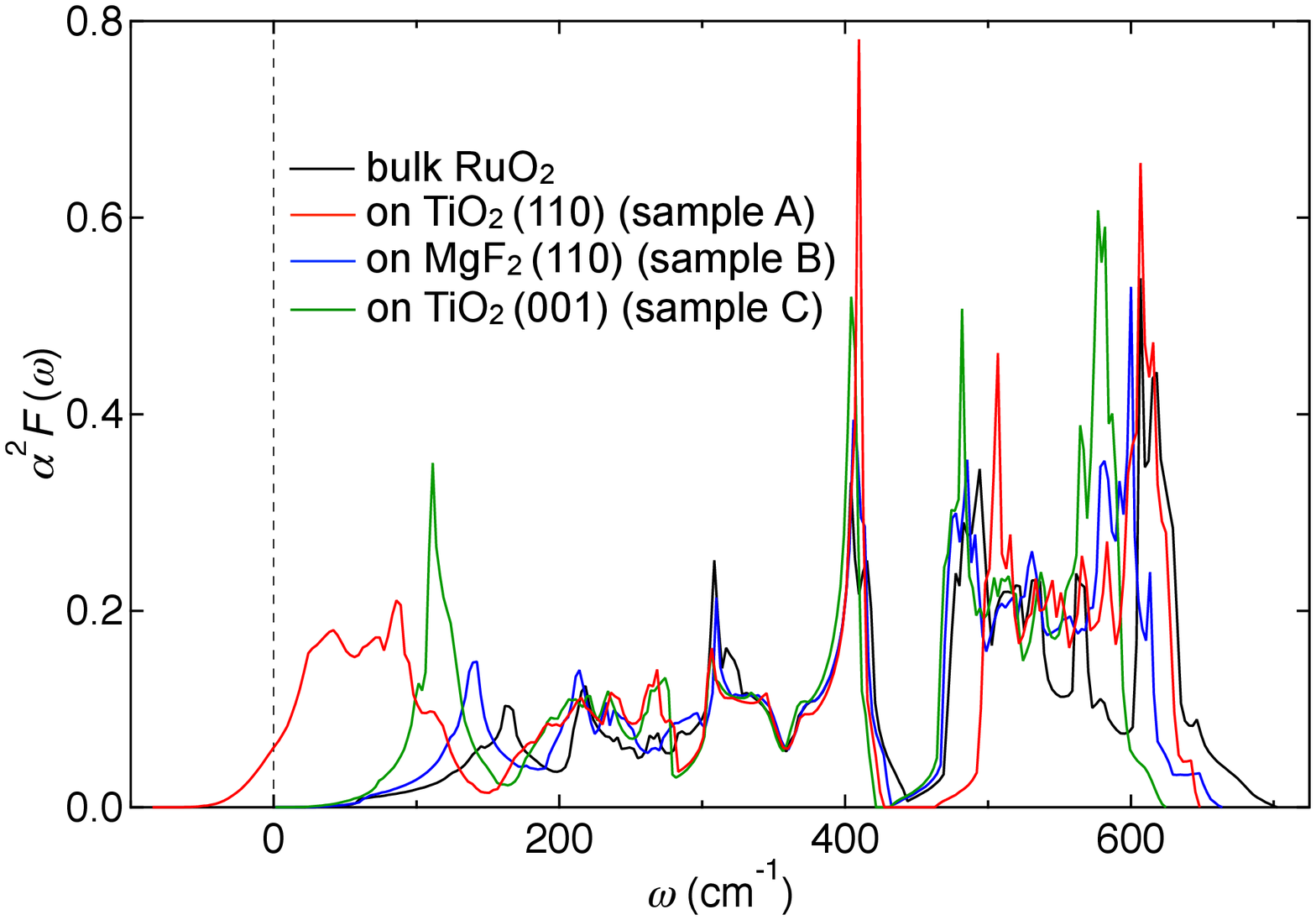}
\caption{
Eliashberg spectral function calculated for the {\RO} bulk and strained thin films (samples A, B, and C).
}
\label{figS3}
\end{center}
\end{figure}

%%%%%%%%%%%%%%%%%%%%%%%%%%%%%%%